\documentclass[usenatbib]{basi}
\usepackage[T1]{fontenc}
\usepackage[british]{babel}
\usepackage[varg]{txfonts}
\usepackage{rotating}
\usepackage{dcolumn}
\begin{document}
\title[Forbush decrease and IP magnetic field]{How are Forbush decreases related with IP magnetic field enhancements ?}
\author[Arunbabu et~al.]%
       {Arunbabu K. ~P.~$^1$,
       P.~ Subramanian$^{1}$ , Sunil Gupta $^2$, H. M. Antia $^2$\\
       $^1$Indian Institute of Science Education and Research, Sai Trinity Building, Pashan, Pune 411 021, India \\
       $^2$Tata Institute of Fundamental Research, Homi Bhabha Road, Mumbai 400 005, India }

\pubyear{2012}
\volume{00}
\pagerange{\pageref{firstpage}--\pageref{lastpage}}

\date{Received --- ; accepted ---}

\maketitle
\label{firstpage}

\begin{abstract}
Cosmic ray Forbush decreases (FDs) are usually thought to be due to
Earth-directed coronal mass ejections (CMEs) from the Sun and their
associated shocks. When CMEs and their shocks reach the Earth, they
cause magnetic field compressions. We seek to understand the relation
between these magnetic field compressions and FDs at rigidities
between 12 and 42 GV using data from the GRAPES-3 instrument at Ooty.
We find that the shapes of the Forbush decrease profiles show a startling similarity
to that of the magnetic field compression in the near-Earth IP medium. We seek to understand the implications of this interesting result.
%
%
\end{abstract}


\section{Introduction}\label{s:intro}
Forbush decreases (FDs) are transient dips in the observed galactic cosmic ray intensity. They can be due to the magnetic field compression of a shock (which acts like an umbrella) (e.g., Wibberenz et al 1998), or due to the  coronal mass ejection behind it (e.g., Subramanian et al 2009). We have examined a large number of FDs observed with the GRAPES-3 muon telescope at Ooty at rigidities ranging from 14 to 24 GV. We have correlated the time profile of these decreases with corresponding magnetic field enhancements in near-Earth interplanetary (IP) space (observed in-situ by spacecraft such as ACE and WIND). We have found that the cosmic ray FDs follow the IP magnetic field enhancements very closely for a number of events. In this paper, we illustrate one such example; the FD observed on 24 November 2001.

We identify the FDs from the cosmic ray intensity data obtained from the $ GRAPES-3 $ muon telescope located at Ooty India. This telescope can get the muon intensity from nine different directions. The data is filtered to eliminate the well known diurnal variation in cosmic ray intensity. The percentage variation of the cosmic ray intensity from the average intensity  for the hourly resolution data is considered for this study. FD events are identified as the sudden decrease in the intensity of cosmic rays and a gradual recovery. We have studied the IP magnetic field variations along with the Forbush decrease. The IP magnetic field data are taken from the in-situ observations from ACE/WIND spacecrafts and are smoothed using the same method we used for the cosmic ray intensity.

\section{Forbush decrease on 24 November 2001}\label{24nov}
The FD observed on 24 November 2001 is associated with a CME first observed in the LASCO field of view on 22 November 2001 at 22:48 UT. The IP shock was observed near the Earth by the WIND spacecraft at 24 November 2001 6:00 UT. The start and end of the magnetic cloud associated with this event were 24 November 2001 17:00 UT and 25 November 2001 13:00 UT respectively. 

\begin{figure}
\centerline{\includegraphics[width=11cm]{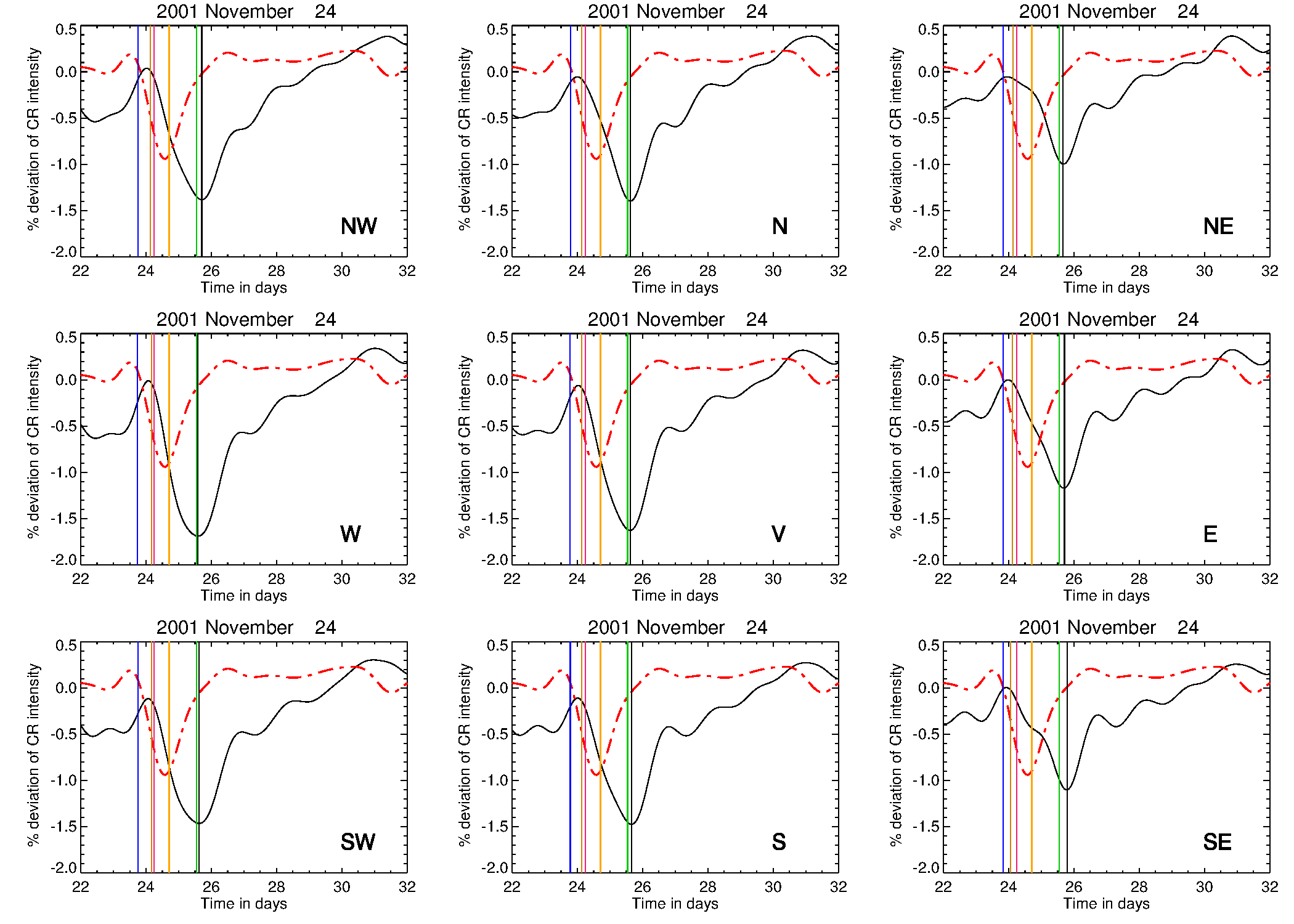}}
\caption{The CR intensity and the IP magnetic field strength during the 24 November 2001 event. The solid black line in each panel shows the percentage deviation of the cosmic ray intensity corresponding to the direction. The red dash-dotted line shows the percentage deviation of magnetic field (100-B).  The vertical lines correspond to the mirror point time (blue), the FD onset time (brown), the FD minimum time (black), the  shock arrival time (magenta), the start (orange) and end (green) time of the magnetic cloud. \label{D24nov}}
\end{figure}

We calculate the percentage variation from the average value of the magnetic field and subtract it from 100. This effectively ``inverts'' the magnetic field enhancement and makes it looks like a decrease, so that we can easily discern its similarity with the FD profile.  The average value for the magnetic field is calculated using a suitable 28 day period, in the same way it is done for the cosmic ray data. In this case, the average is taken from Nov 5 to Dec 2 2001. Since the magnetic field data is smoothed and filtered in the same way as the FD data, we note that the magnetic field enhancement we study is due to the combined effect of the magnetic cloud and the shock. The FD and the magnetic field data for the 24 November 2001 event are shown in Figure~\ref{D24nov}.

\subsection{Correlation} 
\begin{figure}
\centerline{\includegraphics[width=11cm]{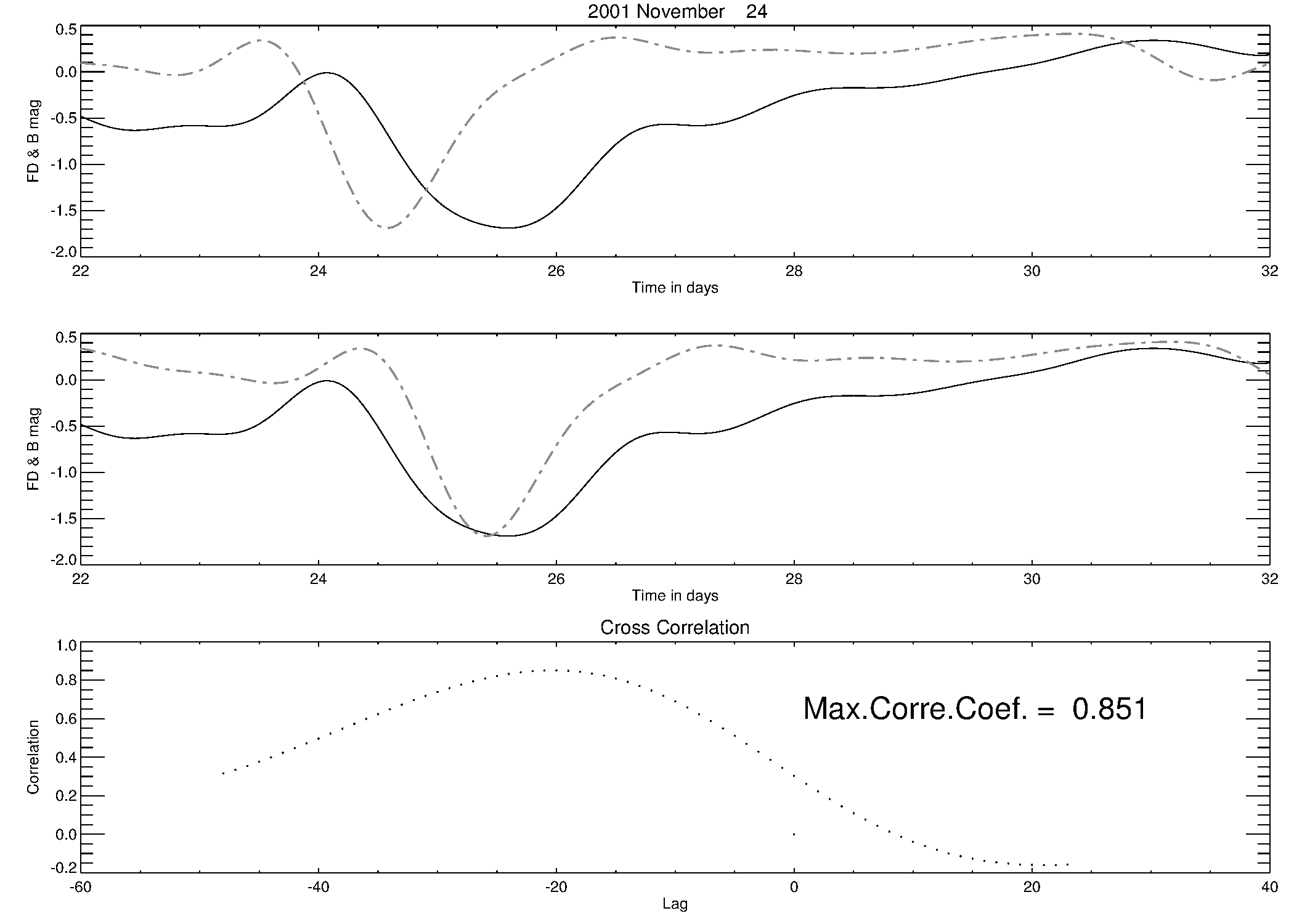}}
\caption{The CR intensity in the vertical direction and the magnetic field strength during the 24 November 2001 event. In the top and middle panels the solid black line shows the percentage deviation of cosmic rays and the dash-dotted line shows the same for the IP magnetic field strength. In the middle panel data points of the IP magnetic field strength are shifted forward in time by the lag corresponding to the maximum correlation. The bottom panel shows the correlation coefficients for different lags.    \label{C24nov}}
\end{figure}

It is evident from Figure~\ref{D24nov} that the FD profile follows the IP magnetic field variations very closely. In order to quantify this, we cross-correlated the FD and IP magnetic field profiles. The result of the cross-correlation procedure is shown in  Figure {\ref{C24nov}}. 
The maximum correlation coefficients and the correlation lag for each direction is given in the Table {\ref{tab}}.

\section{Discussion}
It is clear that cosmic rays follow the magnetic field profile very closely, with a time delay. We investigate this phenomenon by understanding the manner in which the cosmic rays (which are responsible for the FD) interact with the IP magnetic field enhancement. In general, cosmic rays try to penetrate the magnetic field enhancement via gyration and diffusion. For simplicity, we consider these effects only near the Earth, and use the formalism of Kubo \& Shimazu (2010), who model the penetration of cosmic rays into an idealized flux rope structure.

The  classical mirror point, (which is defined as the point where a charged particle will be reflected back) of cosmic rays can be calculated using Equation {\ref{gyr}} ( Kubo \& Shimazu 2010),  
\begin{equation} 
 (\rho_0 \, sin \theta_0  \, sin \phi_0 -  C_{\phi} )^2 + \rho^2 \, ( cos \theta_0 - C_z )^2 - \rho^2 = 0    \label{gyr}
\end{equation}

 The origin of the coordinate system (described in detail in Kubo \& Shimazu 2010) for the quantities described in the equation above is located on the cosmic-ray particle. The quantity $\theta$ is measured from the z-axis, which is parallel to the flux rope axis, while $\phi$ is measured from the r-axis, which points from the center of the flux rope to the particle. We only consider cosmic rays with pitch angles defined by $\theta_0 \, = \, \pi/2 $ and $\phi_0 \, = \, 0  $. The quantity $\rho\, \equiv \, R/R_0 $, where R is the distance from center of flux rope and $R_0$ is its radius. We considered $\rho_0 \, = \, 1 $ in our calculations.

The terms $C_{\phi} $ and $C_z$  are given by Equations \ref{C1} and \ref{C2}.  $f_0 \, = \, R_L / R_0$  where $R_L$ is the larmor radius of the particle.  The constant $a \, = \, 2.40483$, the smallest positive number that gives the zero crossing of the zeroth-order Bessel function of the first kind; i.e., $J_0(a) = 0$. 
$ J_0$ and $ J_1 $ are the usual Bessel functions of the first and second kind respectively.\\

\begin{eqnarray} \label {C}
C_{\phi} & = & \frac{\rho \, J_1 (a\rho) - \rho_0  \, J_1 (a\rho_0 )}{ a\, f_0}  \label {C1} \\  
C_z & = & \frac{ s \,  [ J_0 (a\rho) - J_0 (a\rho_0 )] }{ a\, f_0} \label {C2}
\end{eqnarray}
 
The time delays between the mirror point predicted by Equation~(\ref{gyr}) and the FD onset corresponding to different rigidities in different directions are given in the Table \ref{tab}. The time lags range from 8 to 10 hours. We try to interpret the time delays in the context of cosmic ray diffusion into the flux rope due to MHD turbulence in the shock sheath; i.e., the region between the shock and the CME.


The diffusion of high energy cosmic rays in to the CME is governed by the cross-field diffusion coefficient $D_{\perp}$, which is a function of rigidity of the cosmic ray particle and the level of magnetic turbulence (e.g., Candia \& Roulet 2004).  We define a diffusion length as the perpendicular diffusion coefficient divided by the particle speed, which we take to be equal to the speed of light for these high energy cosmic rays. For the rigidities under consideration here (14--24 GV),  using the turbulence levels discerned for this event from Arunbabu et al (2013), we find that the time lags (of 8 to 10 hours) between the mirror point and the FD onset correspond to around 40 diffusion lengths .

\begin{table}
  \caption{Details of 24 Nov 2001 event.}\label{tab}
  \medskip
  \begin{center}
   \begin{tabular}{|l|c|c|c|c|c|c|}\hline \hline
Direction & Rigidity & Time diff & Time diff & Time diff  & Correlation & Correlation\\
 &  &{ \scriptsize Mirror Point } & { \scriptsize  FD-Onset to} & {\scriptsize  Bmax to  }& lag  &  coefficient\\
&  &{ \scriptsize  to FD-Onset} & { \scriptsize  FD-min} & {\scriptsize   FD-min }&   &  \\
& {\textit{ (GV) }}&{\textit{ (Hrs)}}& {\textit{ (Hrs)}} & {\textit{ (Hrs)} }& {\textit{ (Hrs)} } &  \\
 \hline
NW   &  15.50   &  8.92081    &    37.8792  &      27.0000     &  23    & 0.801777 \\
N      &  18.70   &  8.44080    &    35.6400  &      25.0008     &  23    & 0.822311 \\
NE   &   24.00   &  7.02480   &    36.8808  &      26.0016     &  23     & 0.807123  \\
W     &  14.30   &  10.2648   &    33.9192  &      24.0000     &  20    & 0.850800 \\
V      &  17.20   &  8.72878   &    35.6400  &     25.0008     &  21    &  0.851005  \\
E      &  22.40   &  7.00081   &   38.1192  &      27.0000     &  23     &  0.820339 \\
SW   &  14.40   &  10.0968   &   34.9200  &      25.0008     &  21     & 0.843562 \\
S      &  17.60   &  8.34480   &   36.8808  &      26.0016     &  21     &  0.847526 \\
SE    &  22.40   &  5.56078  &   41.5608  &      29.0016     &  21    & 0.811700  \\ \hline \hline
\end{tabular}\\[5pt]
\end{center}
\end{table}

\section{Conclusion}
We have found that the observed FD time profiles follow those of the IP magnetic field enhancements very closely.
The FD profile lags the IP magnetic field enhancement by -- hours (Correlation lag in Table \ref{tab}). We find that this lag can be explained by a combination of the classical magnetic mirroring effect and cosmic ray diffusion due to MHD turbulence in the shock sheath region. Our current treatment of the time lag takes into account only the cosmic ray pentration near the Earth. However, it is fairly well established that the observed FDs are due to the cumulative effect of cosmic rays penetrating into the CME structure as it propagates from the Sun to the Earth (Subramanian et al 2009; Arun Babu et al 2013). We plan to incorporate the cumulative nature of cosmic ray penetration in future work, so as to explain the observed time lags better.



\section*{Acknowledgements}
K. P. Arunbabu acknowledges support from a Ph.D studentship at IISER Pune. P. Subramanian acknowledges partial support from the CAWSES-II program administered by the Indian Space Research Organization. We thank D. B. Arjunan, A. Jain, the late S. Karthikeyan, K. Manjunath, S. Murugapandian, S. D. Morris, B. Rajesh, B. S. Rao, C. Ravindran, and R. Sureshkumar for their help in the testing, installation, and operating the proportional counters and the associated electronics and during data acquisition. We thank G. P. Francis, I. M. Haroon, V. Jeyakumar, and K.
Ramadass for their help in the fabrication, assembly, and installation of various mechanical components and detectors.

\appendix

\label{lastpage}

\begin{thebibliography}{}
\bibitem[Arunbabu et. al (2013) ]{} Arunbabu, et. al 2013, A \& A , 555, A139 , DOI: 10.1051/0004-6361/201220830
\bibitem[ Candia, J., \& Roulet, E.(2004)]{} Candia, J., \& Roulet, E. 2004, J. Cosmology \& Astropart.\ Phys., 10, 007 
\bibitem[Kubo, Y., \& Shimazu, H. (2010)]{} Kubo, Y., \& Shimazu, H. 2010, Apj, 720, 853
\bibitem[Subramanian et. al (2009) ]{} Subramanian, P. ,et. al 2009, A\& A, 494,1107
\bibitem[Wibberenz, G.,et.al (1998)]{} Wibberenz, G.,et.al 1998, Space Sci.\ Rev., 83, 309
\end{thebibliography}
\end{document}